\documentstyle[epsf,referee]{mn}
\begin{document}

\title{A Detailed Two-Dimensional Stellar Population Study of M~32}
\author[del Burgo et al.]{C. del Burgo,$^{1,2}$,  
R. F. Peletier,$^{3,4}$, A. Vazdekis,$^4$, S. Arribas,$^{1,5}$, and E. Mediavilla,$^{1}$\\
$^1$ Instituto d'Astrof\'\i sica de Canarias, V\'\i a L\'actea, E-38200 La
Laguna, Tenerife, Spain \\
$^2$ Institut d'Astrophysique de Paris, 98bis Boulevard Arago, F-75014 Paris, France \\
$^3$ School of Physics and Astronomy, University of Nottingham, University Park, Nottingham, NG7 2RD, UK \\ 
$^4$ Department of Physics, University of Durham, South Road, Durham, DH1 3LE, UK \\
$^5$ Consejo Superior de Investigaciones Cient\'\i ficas, Spain }

\maketitle

\markboth{del Burgo et al.}{2D Stellar Population Study of M32}

\begin{abstract}

We present Integral Field Spectroscopy of the
9$^{\prime\prime}\times$12$^{\prime\prime}$ circumnuclear region of M~32 obtained with the 2D\_FIS fibre spectrograph installed at the 4.2m William Herschel Telescope. From these spectra line strength maps have been made for
about 20 absorption lines, mostly belonging to the Lick system. We have found
good agreement with long-slit line strength profiles in the literature. We find
no radial gradients in the azimuthally averaged absorption line indices.
We have fitted the mean values of each spectral index and colours from the literature
for the inner regions of M~32 to the models of Vazdekis et al. (1996) and Worthey (1994) finding present data can be well interpreted for a single stellar population of an intermediate age ($\sim$4 Gyr) and a metallicity similar to solar (Z=0.02). 

\end{abstract}


\section{Introduction}

Since M~32 is the brightest elliptical galaxy of the Local Group it has  been
studied extensively as a template for elliptical galaxies further away. 
Stellar population studies, aiming at constraining its formation and evolution
include, for instance, Spinrad \&  Taylor (1969), Faber (1972), O'Connell
(1980),  Burstein et al. (1984), Freedman (1992), Rose (1994) and  Dorman et
al. (1995). The first of those studies showed that the stellar populations of
M~32 are much bluer than those of giant ellipticals. O'Connell (1980) found
that the spectrum of M~32 can be fitted with solar metallicity stars and argued
that the last episode of star formation in this galaxy happened $\sim$ 5 Gyrs
ago. This result has been confirmed by various authors, on the basis of a
number of different stellar population models (e.g. Burstein et al. 1984; Rose
1985, 1994; Bica et al. 1990). This fact distinguishes this galaxy from most other ellipticals, which are found to be much older (e.g., O'Connell 1976,
Aaronson et al. 1978, Tinsley 1980, Rose 1994, Vazdekis et al. 1997, Tantalo et
al. 1998, Peletier et al. 1999, hereafter P99) but also from studies in which is obtained a wide range in age (Gonz\'alez 1993, J\o rgensen 1999, Vazdekis \& Arimoto 1999, Trager et al. 2000). Very recently an extraordinary fit to a high signal-to-noise and high resolution blue spectrum of M~32 has been achieved by Vazdekis \& Arimoto (1999) using a model spectrum  corresponding to a luminosity-weighted mean stellar population of 4~Gyr and solar metallicity selected from the new
single-age, single-metallicity (SSP) spectral library of Vazdekis (1999). The spectrum in the whole optical spectral range of M32, however, is not completely understood at present. When trying to fit all spectral features covering a large spectral range in detail, it seems  that the galaxy cannot be fitted with a single-age, single-metallicity stellar population (see e.g. Burstein et al. (1984), who showed that  H$\beta$ is considerably stronger than expected by their models (see also Faber et al. 1992)). Understanding these small discrepancies will lead to a better understanding of both the stellar  populations of M~32, and of current-day stellar population models.  For that reason the main aim of this paper is to
present a large dataset of well-calibrated data for the central region of M32,
and fit them to current, state-of-the-art  stellar population models. Stellar
population models of old stellar systems have evolved and matured significantly
in the most recent years. Although still following the same concept as the
models of Tinsley \& Gunn (1976), Aaronson et al. (1978), Tinsley (1980),
Arimoto and Yoshii (1986), they have improved considerably, thanks to new sets of isochrone  calculated with improved input physics, including the advanced stages of stellar evolution 
such as the HB and the AGB (e.g., Bertelli et al. 1994;  Dorman et al. 1995) 
and extensive photometric and spectral stellar libraries, either empirical or 
theoretical, which cover wide ranges in atmospheric parameters
(e.g., Worthey et al. 1994, Lejeune et al. 1997). Since most of the
current-day models  (e.g., Worthey 1994, hereafter W94; Vazdekis et al. 1996,
hereafter V96) make use of extensive libraries of empirical stellar spectra  to
predict absorption line-strengths at intermediate resolution (FWHM $\sim$ 9\AA)
in a well-defined system, i.e., the Lick system (Faber et al. 1985; Gorgas et
al. 1993; Worthey et al. 1994; Worthey \& Ottaviani 1997), it is now possible to study not only the ages and metallicities of the galaxies, but also their abundance ratios, which provide information about the physical conditions at the time of formation.

The data that we analyze here come from Integral Field Spectroscopy (IFS) of the central zone of M~32. For the first time for this galaxy 2-dimensional absorption line-strength maps of the inner regions were obtained, 
allowing us to analyze these indices radially and azimuthally. 
Maps were made for most of the Lick indices as well as the Ca{\sc II} IR triplet features of D\'{\i}az et al. (1989). For the interpretation of the data we have used the model of W94, the first one including line-strength predictions for the Lick system, and our own model (V96), which is based on the isochrones of
Bertelli et al. (1994) and which make use of extended empirical stellar input
libraries to calculate the stellar fluxes and the broadband colors instead of
calculating them from the model stellar atmospheres. 
We do not find any noticeable line strength gradients for this galaxy. Although
this is not surprising, given the fact that there is no evidence for colour gradients 
(Peletier 1993, Lauer 1998) in the inner regions, and for line strength 
gradients in previous work (e.g. Trager et al. 1998, hereafter T98, Gonz\'alez 1993), there are very few
other galaxies where colour and line strength profiles are so flat. The aim
of this paper has also been to address this problem by investigating gradients 
in more line indices than before, using higher-quality data.

The organization of this paper is as follows.
In \S~2 we present these
observations and the data reduction. In \S~3 we fit the data to predictions from
the stellar population synthesis model. The results are shown in \S~4 while a
discussion and our conclusions are presented in \S~5 and \S~6, respectively.


\section{Observations and Data Reduction}


The data were obtained on February 15 and 16 1997 at the Observatorio del Roque de los Muchachos on the Island of La Palma. We used the 2D-FIS (Two-dimensional Fiber ISIS System), which linked the f/11 Cassegrain focus of the 4.2m William
Herschel Telescope (WHT) and the ISIS (Carter et al. 1993) double 
spectrograph. A detailed description is provided by Garc\'\i a et al. (1994). 
The main characteristics of our setup are given in P99. In the blue arm, the
R300 grating was used, giving us a reciprocal dispersion of 1.55 \AA\
pixel$^{-1}$ (96 km s$^{-1}$), a spectral range between 4100 and 5650 \AA,
covering most of the  lines of the Lick system, including H$\gamma$  (Worthey
\& Ottaviani  1997), and a resolution of 2 pixels. In the red, we centred our 
spectrum on the Ca II IR triplet. Here we had a pixel size of 28 km s$^{-1}$ 
(resolution 2 pixels) and a spectral range from 8200 to 9000 \AA. On  both arms
1024 $\times$ 1024 thinned TEK CCDs were used. A dichroic,  centred at 6100 \AA,
was used, and in the red arm a GG495 order sorting  filter was placed as well.
Weather conditions were photometric, with  seeing of about 1 arcsec. The galaxy
was observed  each night for 1800  seconds. Also, each night 20 out-of-focus
standard stars of the Lick  system were observed using the same setup, which were so
far out of focus that all fibers were exposed.

\subsection{Data Reduction}

For a full description of the data reduction the reader is referred to P99,
describing the IFS observations  of the standard early-type galaxies NGC~3379,
NGC~4472 and NGC~4594, which were obtained during the same observing run. Here,
we concentrate on the conversions applied to the data in order to translate our
M~32 index measurements to the Lick and D\'\i az et al. (1989) systems. It is
important to apply them before attempting any comparison with the model
predictions (see Worthey \& Ottaviani 1997 for a review and Vazdekis et al.
1997 for a practical application). The first step is to investigate how the
index measurements vary when smoothing our spectral resolution (192~km~s$^{-1}$
and 56 km~s$^{-1}$ for the blue and red ranges, respectively) to match those of
mentioned systems. A velocity dispersion correction curve was obtained by first
broadening our 20 stellar spectra with different Gaussians and then measuring
all the spectral indices. For each standard star we averaged the values that we
measured on each aperture (fibre image). Since no important differences were
found among curves obtained from the different stars of type G, K and early M
we calculated a mean velocity dispersion correction curve. The standard stars
also helped us to estimate the mean velocity dispersion of M~32 in the central
9$^{\prime\prime}\times$12$^{\prime\prime}$ (84~km~s$^{-1}$ for the blue
spectrum and 87~km~s$^{-1}$ for the red). Note that since this dispersion is
quite low compared with that of the Lick system (larger than 200~km~s$^{-1}$)
the velocity dispersion corrections to be applied are much smaller than those
applied for the giant ellipticals studied in P99 (with  velocity dispersions
$\sim$200~km~s$^{-1}$). Table \ref{corrdisp} lists these corrections (via
parameter VDCorr) for M~32. The corrected values for each index are obtained by
dividing the observed value by VDCorr, except for Mg1 and Mg2 for which 
they have to be 
added. Note that the corrections for the Ca{\sc II} triplet features are very
similar to that for Fe5270, larger than those for H$\beta$ and lower than those
for Ca4455. 

\begin{table}
\caption[]{Velocity dispersion correction coefficients for M32}
 \footnotesize
\label{corrdisp}
\[
\begin{array}{p{.10\linewidth}rr}
\hline
    \hline
\noalign{\smallskip}
Index & VDCorr \\
\noalign{\smallskip}
\hline
\noalign{\smallskip}
H$\gamma$a  &	 1.007  \\
H$\gamma$f  &	 1.042  \\
Ca4227      &	 0.9447 \\
G4300	    &	 0.9285 \\
Fe4383      &	 0.9996 \\
Ca4455      &	 0.9073 \\
Fe4531      &	 0.9626 \\
C$_{2}$4668 &	 1.001  \\
H$\beta$    &	 0.9966 \\
Fe5015      &	 0.9563 \\
Mg1         &    -0.0001467 \\
Mg2         &    -0.0003682 \\
Mgb	    &	 0.9711 \\
Fe5270      &	 0.9628 \\
Fe5335      &	 0.9357 \\
Fe5406      &	 0.9409 \\
CaII2	    &	 0.9759 \\
CaII3	    &	 0.9660 \\
\noalign{\smallskip}
\hline  							       
\end{array}
\]
\end{table}

The next step is to place our spectral indices on the Lick/IDS and  D\'\i az 
et al. (1989) systems by correcting for the fact that our 
instrumental response curve is not the same as the one used for those systems.
Linear relations which  allow us 
to carry out this transformation were obtained by comparing the average values
of the index measurements for the  out-of-focus  stars with the values
tabulated by the above mentioned authors for the same  stars. For  the Ca{\sc II} IR
features an offset was sufficient. However, for the  blue features our response
curve was considerably different from that of the Lick Intermediate Dispersion
Spectrograph (IDS), used to define the  Lick system, and a
linear conversion was required. Table 3 of P99 lists the coefficients for the
equation  O = a$\pm\Delta$a + (b$\pm\Delta$b) C, where O and C are the
observed  and calibrated indices, respectively. The errors $\pm\Delta$a and
$\pm\Delta$b are derived from the dispersion in the values of the stars. The
above-mentioned Table also lists the RMS dispersion in this relation for a line
strength measurement of an individual star and the RMS dispersion from aperture
to aperture.

Since M~32 was observed at high airmasses (1.9 and 1.7 in the first and  second
nights, respectively), a correction for the effects of differential atmospheric refraction (DAR) was necessary. Atmospheric refraction has the effect that monochromatic images of M32 in the focal plane of the WHT are shifted depending on their wavelength. To correct for this effect we used an algorithm which takes into account the dependency of the refraction index on the wavelength (Allen 1976), and determined the spatial shifts on the sky between the maxima of the continuum at the central wavelengths of the spectral indices (Arribas et al. 1999).

\subsection{The index maps and resulting measurements}

\begin{figure}
\mbox{\epsfxsize=16cm  \epsfbox{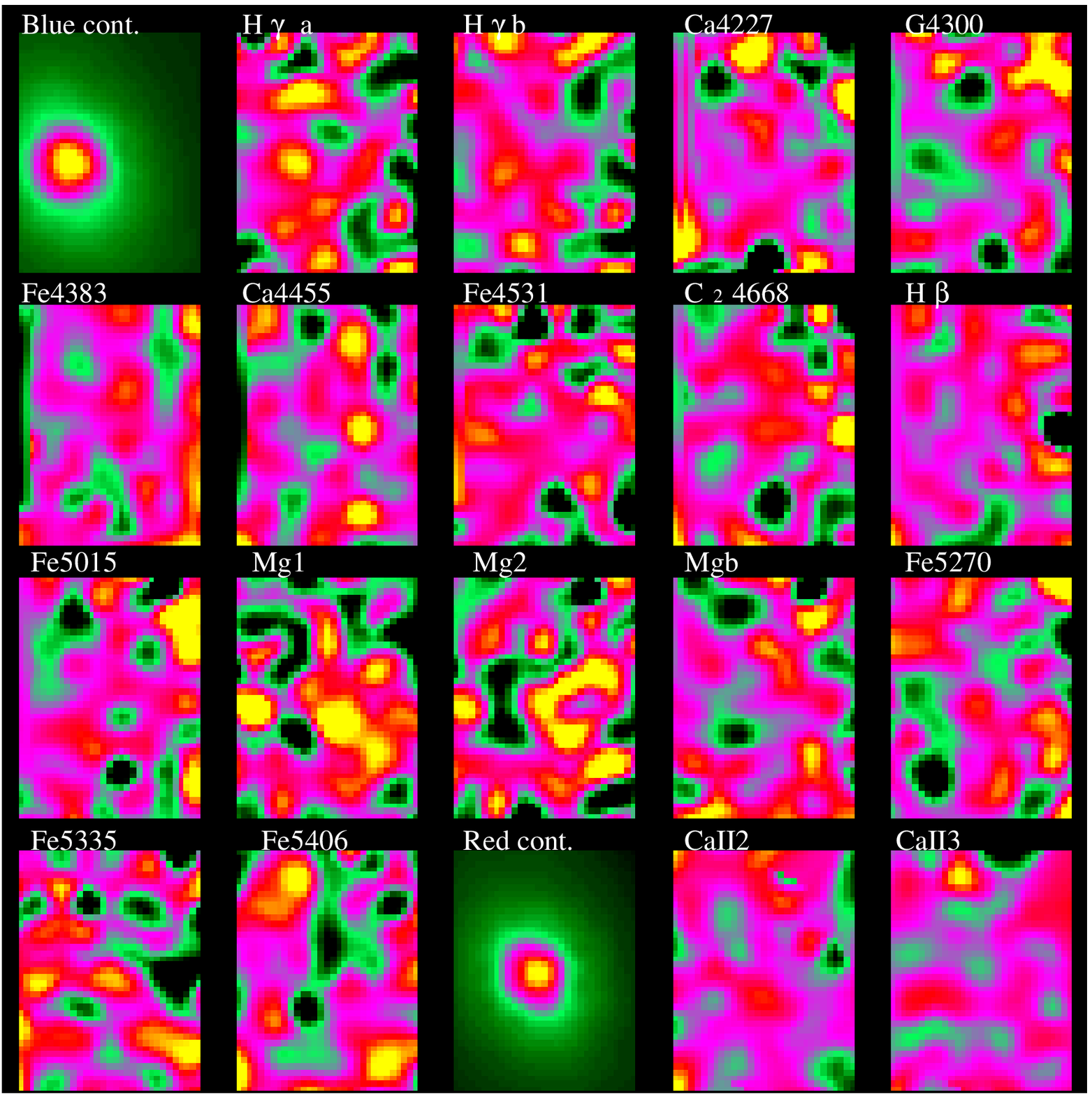}}
\caption{Maps of the most important indices in the blue and red for M~32. The
maps have been scaled in a such way that pixels that are more than 2$\sigma$
fainter than the mean value of a map are black, while those more than 2$\sigma$
brighter are white. If there is a very little structure in a map, it look
noisier than in a case where the dynamic range in the map is large. The size of
the map is 8.2$^{\prime\prime}\times$11$^{\prime\prime}$.
\label{Figure1}}
\end{figure}

Once all these corrections were carried out line-strength maps (see Fig.
\ref{Figure1}) were obtained for the central zone of M~32 following the
procedure described in P99. Note that the small structures present in the
spectral index maps are mainly caused by the noise. 

\begin{figure}
\mbox{\epsfxsize=16cm  \epsfbox{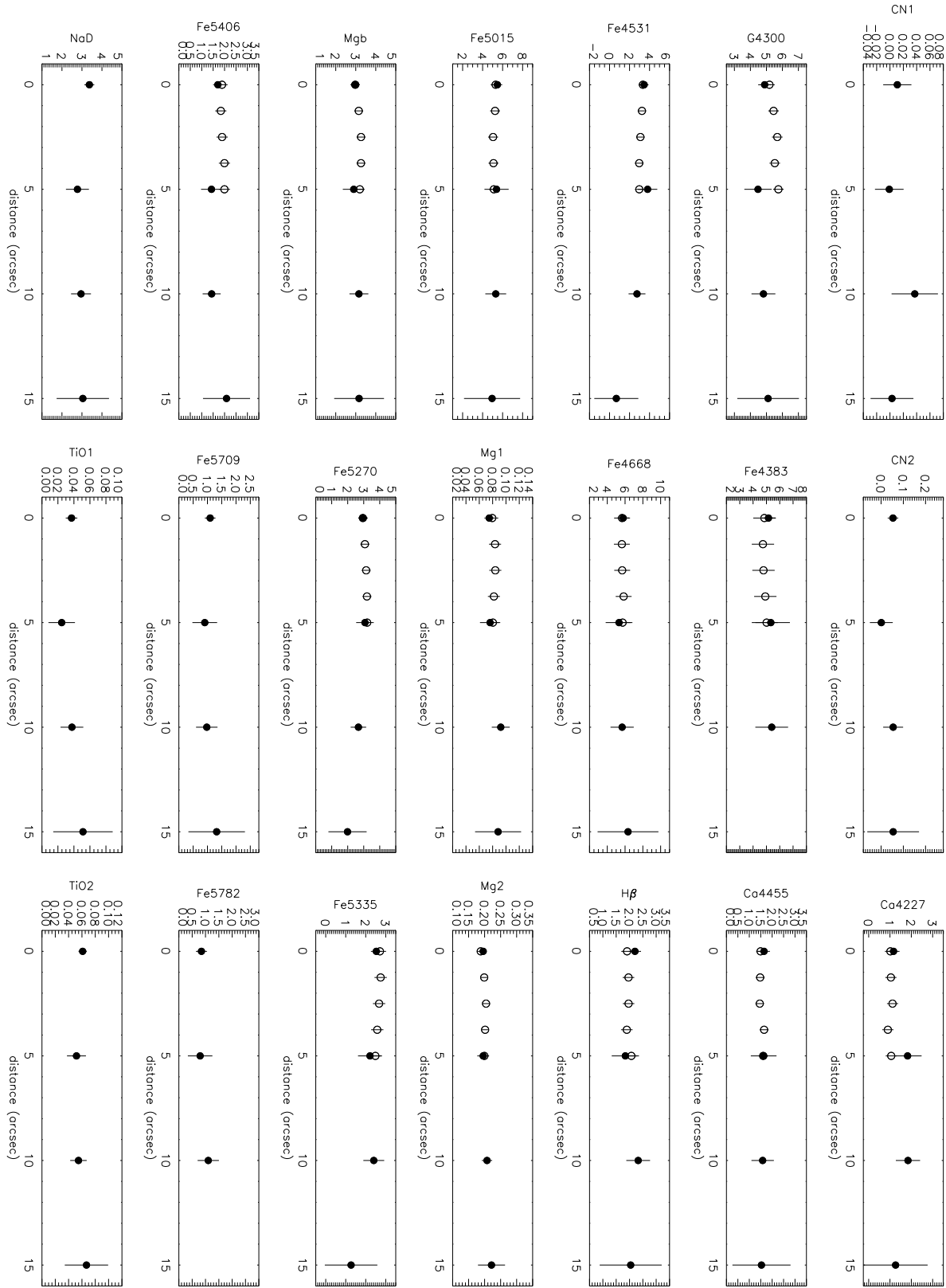}}
\caption[Figure2.ps]{Spectral indices as a function of radius for 2D\_FIS
data (open circles) and data of T98 (filled circles). 
\label{Figure2}}
\end{figure}

Next, the line indices were azimuthally averaged, and radial gradients were
determined in the following way:

\begin{itemize} \item[1.] We fitted five ellipses (with semi-major axes of
1.25$^{\prime\prime}$, 2.5$^{\prime\prime}$, 3.75$^{\prime\prime}$,
5.0$^{\prime\prime}$, 6.25$^{\prime\prime}$) to the isophotes of the blue and
red continuum maps obtained by integrating over the whole spectral range. We
used the  {\tt ellipse} task of the {\tt galphot} package (see J\o rgensen et
al. 1992).

\item[2.] We obtained the mean values for each spectral index in 
the central ellipse and in rings bound by ellipses with semi-major
axis (sma) of 1.25$^{\prime\prime}$,
2.5$^{\prime\prime}$,
3.75$^{\prime\prime}$,
5.0$^{\prime\prime}$ and 6.25$^{\prime\prime}$ (with the centres corrected for
differential refraction) at each wavelength. Note that the ellipses
corresponding to the red continuum map were used only for the Ca{\sc II}
triplet features.

\item[3.] Each mean spectral index was plotted vs. the sma of these rings 
(see Fig. \ref{Figure2}) and we determined a straight line of the form: \begin{equation}
index=(i_0 \pm \Delta i_0) + (m \pm \Delta m) \times r(^{\prime\prime})
\end{equation} using a least squares fit from all the positions considered (N=5). The fits are shown in Table~\ref{FSL}. These fits take into account the errors in the spectral index measurements (shown in Table \ref{MODall}), which are determined from the
fibre-fibre mean dispersion  (see P99 for details). The Chi-square test ($\chi^2 = \sum_{j=1}^{N} \left(\frac{index_j - i_0  - m \times  r_j}{error_j}\right)^2$, where error$_j$ stands for the RMS in index$_j$) inform us about the quality of the fit. Data indicated by T98 come from fits to longslit spectra of T98 between the center and 15$''$ (N=4 in this case).
\end{itemize}

A very important result emerges when looking at Table~\ref{FSL} and Fig.
\ref{Figure2} is the fact that no gradients have been found. For none 
of the indices the variations in the sampled area  are 
significant,
because they are within the error bars. The absence of gradients allow us to
obtain a mean value for each index in the field covered by 2D\_FIS. Table
\ref{OVal} lists these mean values and their RMS for the first and the second
night.

        \begin{table}
        \caption[]{Straight line fits for the indices of M32}
        \footnotesize
        \label{FSL}
        \[
        \begin{array}{p{.08\linewidth}p{.08\linewidth}rrrrrrrr}
        \hline
        \hline
        \noalign{\smallskip}
         Index & Units & i_0 & \Delta i_0 & m & \Delta m & \chi^2 & 
 Source \\
        \noalign{\smallskip}
        \hline
        \noalign{\smallskip}
        \ CN1     & mag & 0.01  & 0.03  &  0.000 & 0.002 &  0.9112  & 
 T98^{\mathrm{a}} \\
        \ CN2     & mag & 0.05  & 0.04  &  0.000 & 0.004 &  0.8747  & 
 T98 \\
        \ Ca4227  & \AA & 1.0   & 0.2   & -0.01  & 0.06  &  0.3782 & 
 N1^{\mathrm{b}} \\
        \         &  & 1.1   & 0.2   &  0.00  & 0.06  &  0.0754  & 
 N2^{\mathrm{c}}\\
        \         &  & 0.9   & 0.5   &  0.06  & 0.05  &  0.5742  & T98 
 \\
        \ G4300   & \AA & 5.3   & 0.3   &  0.09  & 0.08  &  0.4552 &  N1 \\
        \         &  & 5.5   & 0.3   &  0.00  & 0.08  &  0.3991  & N2 
 \\
        \         &  & 4.9   & 0.3   & -0.01  & 0.05  &  0.2670  & T98 
 \\
        \ Fe4383  & \AA & 4.8   & 0.6   &  0.0   & 0.2   &  0.0313  & 
 N1 \\
        \         &  & 5.0   & 0.6   & -0.1   & 0.2   &  0.0111 & N2 
 \\
        \         &  & 4.9   & 0.7   &  0.01  & 0.07  &  0.2060  & T98 
 \\
        \ Ca4455  & \AA & 1.5   & 0.1   &  0.03  & 0.05  &  0.3601  & 
 N1 \\
        \         &  & 1.6   & 0.1   &  0.01  & 0.05  &  0.2185 & N2 
 \\
        \         &  & 1.7   & 0.4   &  0.01  & 0.05  &  0.0009 & T98 
 \\
        \ Fe4531  & \AA & 3.3   & 0.3   & -0.1   & 0.1   &  0.0400 & 
 N1 \\
        \         &  & 2.7   & 0.3   &  0.1   & 0.1   &  0.2424 & N2 
 \\
        \         &  & 4.0   & 0.7   & -0.09  & 0.08  &  1.0607 & T98 
 \\
        \ C$_{2}$4668  & \AA & 5.6   & 0.7   &  0.0   & 0.2   &  0.0186  & N1 \\
        \         &  & 5.7   & 0.7   &  0.0   & 0.2   &  0.0489  & N2 
 \\
        \         &  & 6           & 1     &  0.0   & 0.1   &  0.1118 & T98 \\
        \ H$\beta$ & \AA & 1.9   & 0.2   &  0.02  & 0.05  &  0.2517 & 
 N1 \\
        \         &  & 2.0   & 0.2   & -0.01  & 0.05  &  0.0340 & N2 
 \\
        \         &  & 2.2   & 0.4   &  0.00  & 0.04  &  0.5654 & T98 
 \\
        \ Fe5015  & \AA & 5.3   & 0.4   &  0.0   & 0.1   &  0.0845 & 
 N1 \\
        \         &  & 5.2   & 0.4   &  0.0   & 0.1   &  0.1414 & N2 
 \\
        \         &  & 5.6   & 0.8   & -0.02  & 0.09  &  0.0082 & T98 
 \\
        \ Mg1     & mag & 0.082 & 0.007 &  0.000 & 0.002 &  0.1956 & 
 N1 \\
        \         &  & 0.074 & 0.007 &  0.001 & 0.002 &  0.1708 & N2 
 \\
        \         &  & 0.07  & 0.01  &  0.001 & 0.001 &  0.2448 & T98 
 \\
        \ Mg2     & mag & 0.193 & 0.008 &  0.002 & 0.003 &  0.9094 & 
 N1 \\
        \         &  & 0.191 & 0.008 &  0.002 & 0.003 &  0.1461 & N2 
 \\
        \         &  & 0.19  & 0.01  &  0.001 & 0.002 &  0.1325 & T98 
 \\
        \ Mgb     & \AA & 3.1   & 0.1   &  0.04  & 0.05  &  0.5168 & 
 N1 \\
        \         &  & 3.0   & 0.1   &  0.01  & 0.05  &  0.2481 & N2 
 \\
        \         &  & 3.2   & 0.5   & -0.04  & 0.05  &  0.4017 & T98 
 \\
        \ Fe5270  & \AA & 3.0   & 0.2   &  0.05  & 0.07  &  0.0620 & 
 N1 \\
        \         &  & 3.0   & 0.2   &  0.00  & 0.07  &  0.0362 & N2 
 \\
        \         &  & 3.1   & 0.2   & -0.03  & 0.03  &  0.5262 & T98 
 \\
        \ Fe5335  & \AA & 2.8   & 0.2   & -0.05  & 0.08  &  0.0720 & 
 N1 \\
        \         &  & 2.5   & 0.2   & -0.01  & 0.08  &  0.1154 & N2 
 \\
        \         &  & 2.7   & 0.4   & -0.04  & 0.05  &  0.5526 & T98 
 \\
        \ Fe5406  & \AA & 1.8   & 0.2   &  0.03  & 0.06  &  0.1025  & 
 N1 \\
        \         &  & 1.8   & 0.2   & -0.02  & 0.06  &  0.1994 & N2 
 \\
        \         &  & 1.8   & 0.3   & -0.02  & 0.04  &  0.5998 & T98 
 \\
        \ Fe5709  & \AA & 1.1   & 0.3   & -0.01  & 0.04  &  0.2397 & 
 T98 \\
        \ Fe5782  & \AA & 0.7   & 0.3   &  0.02  & 0.04  &  0.1213 & 
 T98 \\
        \ NaD     & \AA & 3.5   & 0.4   & -0.04  & 0.05  &  0.4664 & 
 T98 \\
        \ TiO1    & mag & 0.04  & 0.01  &  0.000 & 0.001 &  0.6614 & 
 T98 \\
        \ TiO2    & mag & 0.06  & 0.01  &  0.000 & 0.001 &  0.3807 & 
 T98 \\
        \ H$\gamma$a & \AA & -3.7     &   0.6   &  -0.1    &  0.2   &  0.0038 &  N1        \\
        \         &    & -4.0     &   0.6   &   0.0    &  0.2   &  0.0386 &  N2  \\
        \ H$\gamma$f & \AA & -0.6     &   0.2   &  -0.04   &  0.07  &  0.0728 &  N1        \\
        \         &    & -0.6     &   0.2   &  -0.05   &  0.07  &  0.1301 &  N2 \\
        \ Ca1     & \AA   & 1.9        &   0.6   &  -0.1    &  0.2   &  0.1006 &  N1 \\
        \         &    & 1.9    &   0.6   &   0.0    &  0.2   &  0.0518 &  N2 \\
        \ Ca2     & \AA   & 4.0        &   0.6   &   0.0    &  0.2   &  0.0580 &  N1 \\
        \         &    & 4.4    &   0.6   &   0.0    &  0.2   &  0.0748 &  N2 \\
        \ Ca3     & \AA   & 3.4        &   0.6   &   0.0    &  0.2   &  0.0415 &  N1 \\
        \         &    & 3.1    &   0.6   &   0.2    &  0.2   &  0.2717 &  N2 \\
        \ Ca2+3   & \AA   & 7.4        &   0.8   &  -0.1    &  0.3   &  0.0388 &  N1 \\
        \         &    & 7.6    &   0.8   &   0.2    &  0.3   &  0.3022 &  N2 \\
        \noalign{\smallskip}
         \hline                                                                 
         \end{array}
         \]
	 \vfil\eject
	 \end{table}

	\addtocounter{table}{-1}
	\begin{table}
	\caption[]{Straight line fits - continued}
        \[
        \begin{array}{p{.08\linewidth}p{.08\linewidth}rrrrrrrr}
        \hline
        \hline
        \noalign{\smallskip}
         Index & Units & i_0 & \Delta i_0 & m & \Delta m & \chi^2 & 
 Source \\
        \noalign{\smallskip}
        \hline
        \noalign{\smallskip}
        \ CaAZ1   & \AA   & 1.7        &   0.3   &  -0.03   &  0.09  &  0.2363 &  N1 \\
        \         &    & 1.6    &   0.3   &  -0.03   &  0.09  &  0.1611 &  N2 \\
        \ CaAZ2   & \AA   & 3.6        &   0.3   &  -0.03   &  0.01  &  0.2362 &  N1 \\
        \         &    & 3.8    &   0.3   &  -0.11   &  0.01  &  0.4563 &  N2 \\
        \ CaAZ3   & \AA   & 2.9        &   0.3   &   0.01   &  0.09  &  0.2146 &  N1 \\
        \          &    & 2.9   &   0.3   &   0.02   &  0.09  &  0.4930 &  N2 \\
        \ CaAD1   & \AA   & 3.7        &   0.3   &   0.0    &  0.1   &  0.1379 &  N1 \\
        \         &    & 3.9    &   0.3   &   0.0    &  0.1   &  0.2184 &  N2 \\
        \ CaAD2   & \AA   & 3.0        &   0.3   &   0.0    &  0.1   &  0.2490 &  N1 \\
        \         &    & 2.9    &   0.3   &   0.0    &  0.1   &  0.4718 &  N2 \\
        \ CaTP1   & \AA   & 1.8        &   0.3   &   0.0    &  0.1   &  0.1144 &  N1 \\
        \         &    & 1.9    &   0.3   &   0.0    &  0.1   &  0.0592 &  N2 \\
        \ CaTP2   & \AA & 3.9  &   0.3   &   0.0    &  0.1   &  0.0794 &  N1 \\
        \         &    & 4.2    &   0.3   &  -0.1    &  0.1   &  0.3166 &  N2 \\
        \ CaTP3   & \AA   & 2.7        &   0.3   &   0.0    &  0.1   &  0.2064 &  N1 \\
        \         &    & 2.6    &   0.3   &   0.1    &  0.1   &  0.1931 &  N2 \\
        \noalign{\smallskip}
         \hline                                                                 
         \end{array}
         \]
   \begin{list}{}{}
   \item[$^{\mathrm{a}}$] long-slit between 0$^{\prime\prime}$ and 15$^{\prime\prime}$ of distance from the centre (T98).
   \item[$^{\mathrm{b}}$] 2D$\_$FIS data from the 15 February, Night 1.
   \item[$^{\mathrm{c}}$] 2D$\_$FIS data from the 16 February, Night 2.
 \end{list}
         \end{table}

\begin{table}
\caption[]{Observed average spectral indices}
\footnotesize
\label{OVal}
\[
\begin{array}{p{.07\linewidth}p{.07\linewidth}rrrrrr}
\hline
\hline
\noalign{\smallskip}
~ & ~ & \multicolumn{2}{c}{Night 1} & \multicolumn{2}{c}{Night 2} \\
Index & Units & mean & error & mean & error \\
\noalign{\smallskip}
\hline
\noalign{\smallskip}
H$\gamma$a & \AA & -4.1  & 0.5  & -3.9  & 0.5  \\
H$\gamma$f & \AA & -0.7  & 0.3  & -0.7  & 0.3  \\
Ca4227     & \AA &  1.0  & 0.2  & 1.0  & 0.2  \\
G4300     & \AA &  5.7  & 0.5  & 5.5  & 0.5  \\
Fe4383     & \AA &  5.0  & 0.6  & 4.7  & 0.6  \\
Ca4455     & \AA &  1.5  & 0.3  & 1.7  & 0.2  \\
Fe4531     & \AA &  3.0  & 0.5  & 2.9  & 0.8  \\
C$_{2}$4668     & \AA &  5.7  & 0.8  & 5.8  & 0.6  \\
H$\beta$   & \AA &  2.0  & 0.3  & 2.0  & 0.2  \\
Fe5015     & \AA &  5.2  & 0.5  & 5.1  & 0.5  \\
Mg1       & mag  &  0.08 & 0.01 & 0.08 & 0.01  \\
Mg2       & mag  &  0.20 & 0.01 & 0.20 & 0.01 \\
Mgb       & \AA &  3.2  & 0.3  & 3.0  & 0.2  \\
Fe5270     & \AA &  3.2  & 0.3  & 3.0  & 0.2  \\
Fe5335     & \AA &  2.6  & 0.3  & 2.5  & 0.2  \\
Fe5406    & \AA &  1.9  & 0.3  & 1.8  & 0.3  \\
CaII2     & \AA &  3.8  & 0.6  & 3.8  & 0.6  \\
CaII3     & \AA &  3.4  & 0.7  & 3.4  & 0.7  \\
\noalign{\smallskip}
\hline                                                                 
\end{array}
\]
\end{table}

\subsection{Comparison with previous results}

Fig. \ref{Figure2} shows our index measurements compared to those of T98,
showing very good agreement.  Note that despite the fact that their results
were achieved with  long-slit spectroscopy we can compare with our 2D\_FIS
field,  since no gradients were found and therefore the indices do not depend on
the slit position. The only significant  difference is obtained for the G
band (which is $\sim$15 $\%$ or 1.8 $\sigma$ larger in our case).         

\section{The stellar population analysis}

\subsection{Observational constrains}

The next step was to analyze the stellar populations. We decided
to add to the data, in addition to the line  strengths,
also a set of optical and near-infrared colours obtained by Peletier
(1993), mainly to constrain the shape of the continuum and to include spectral indices at other different spectral regions for which different stars contribute in a different way. 
We averaged the colours in the same region of the galaxy, and corrected
for Galactic extinction using the DIRBE-maps analyzed by Schlegel et al.
(1998), and the Galactic extinction law given by  Rieke \& Lebofsky (1985). In
Table \ref{MODall}, the colours that we used are given together with the line indices.  We used various age indicators,  such as H$\beta$ (W94), and H$\gamma$A and H$\gamma$F (Worthey  \& Ottaviani 1997). These indices when plotted against strong metallicity indicators like the iron or magnesium dominated lines are able to constrain reasonably well both the age and the metallicity of the stellar population  (e.g., W94; Vazdekis et al. 1997). When these metallicity indicators are plotted against each other they show that giant ellipticals  have [Mg/Fe]$>$0.0 (e.g., Peletier 1989; Worthey et al. 1992).We also observed the Ca4227 index, which was found to be much lower than expected in a number of early-type galaxies by  Vazdekis et al. (1997) and P99. Index-index diagrams allow
us to note the very strong CN absorption of the  metal-rich globular clusters
(e.g.: Rose 1994; Vazdekis 1999a). One of the metal indicators is C$_2$4668, shown by Tripicco \& Bell (1995) to be an excellent indicator for the global metallicity.
This index was found very useful to analyze
the Fornax early-type galaxies (Kuntschner \& Davies 1998). We have also
included a number of indicators which can help in constraining the dwarf/giant
ratios and the IMF such as the Ca{\sc II} triplet features (D\'{\i}az et al.
1989) and TiO$_{1}$ (see  V96 (Table~2) and Vazdekis et al. 1997 (Fig.~10)).
Vazdekis (1999b) showed that no powerful IMF indicators are present in the
range  $\lambda\lambda$ (4000-5500\AA) (Mg$_{1}$ being the most sensitive of
these).

\subsection{Models} 

For the stellar population analysis we decided to use the models of W94 and
V96. These two models predict single-age single-metallicity stellar populations
for intermediate and old stellar populations. The models of V96 have recently
been updated by implementing the new metallic dependent  empirical relations of
Alonso et al. (1999) for giants with temperatures larger than $\sim$3500~K for
all metallicities, and by applying a semi-empirical approach for the coolest
dwarfs and giants on the basis of the empirical color-T$_{eff}$ relations of
Lejeune et al. (1997) and Lejeune et al. (1998) for solar metallicity, and the 
stellar model atmospheres of Hauschildt et al. (1999).  These models are (see
{\tt http://star-www.dur.ac.uk/$\sim$vazdekis} for the latest version)  now
more accurate and cover a larger range of ages and metallicities. The
differences with V96 in general are small, especially for solar metallicity;
only the IR colors vary slightly. W94 uses a Salpeter (1955) IMF, while V96
provides models with 2 different IMFs: the Unimodal, which is a power-law, with
the slope as  a free parameter (where 1.3 corresponds to Salpeter), and a
Bimodal IMF, where the number of stars of mass $<$0.6 M$_{\odot}$ has been
truncated. These models all include the optical spectral indices of the
Lick/IDS system. The V96 model also predicts the Ca{\sc II} triplet  features
on the system of D\'{\i}az et al. (1989).        

\subsection{The fits} 

\begin{figure}
\mbox{\epsfysize=19cm  \epsfbox{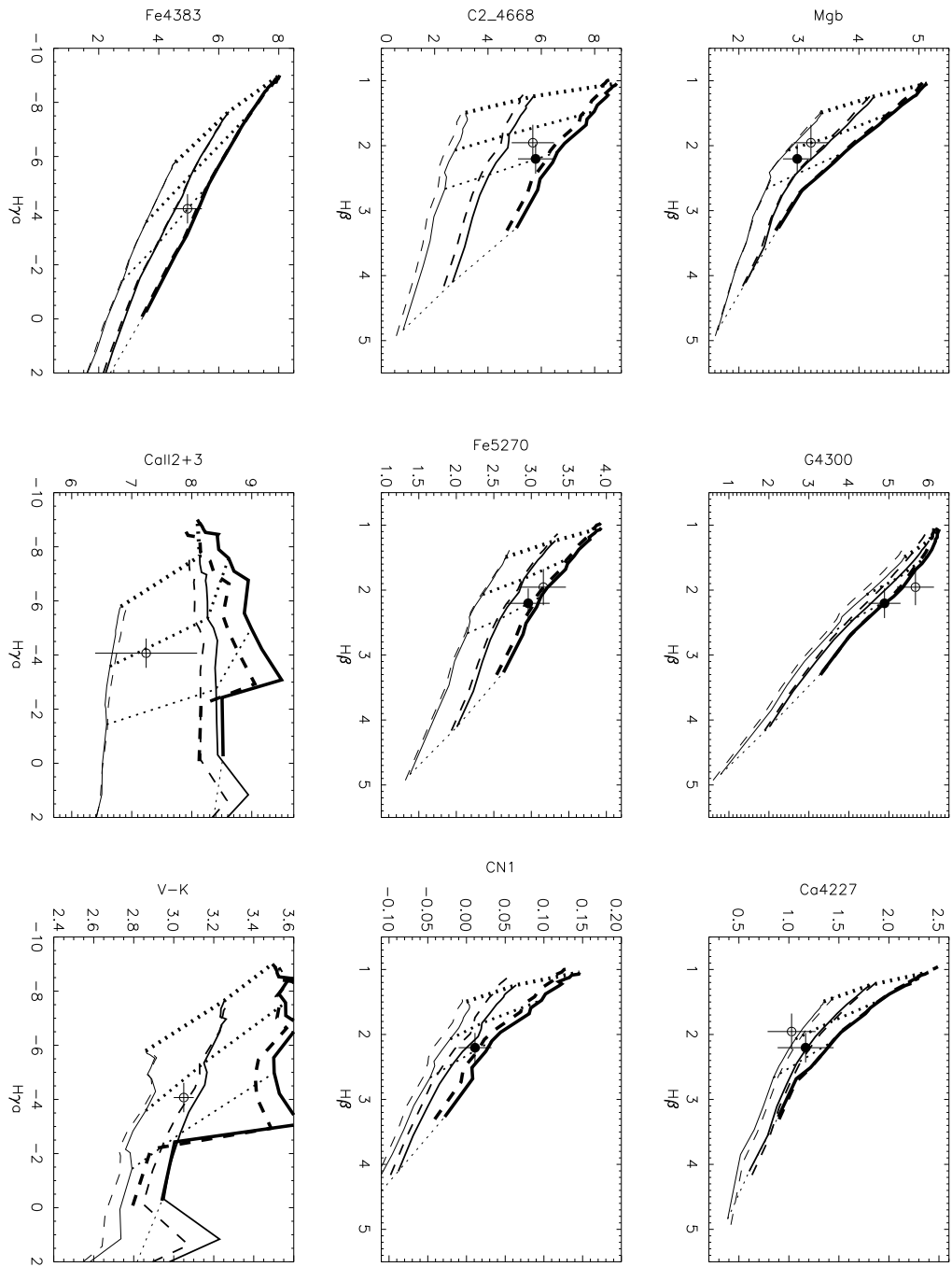}}
\caption[Figure3.ps]{Spectral index-index diagrams. Data from here (open circles) and from T98 (filled circles) are plotted. The
bimodal models with IMF slope of 1.35 (continuous line) and 2.35 (dashed line)
of Vazdekis et al. (1996) are plotted for metallicites Z=0.008 (thin line),
0.02 (intermediate line) and 0.05 (thick line). Ages correspond to  1.0, 2.5, 6.3 and
17.4 Gyrs, respectively, from  the thinnest to thickest dotted line.
\label{Figure3}}
\end{figure}

\begin{figure}
\mbox{\epsfysize=22cm  \epsfbox{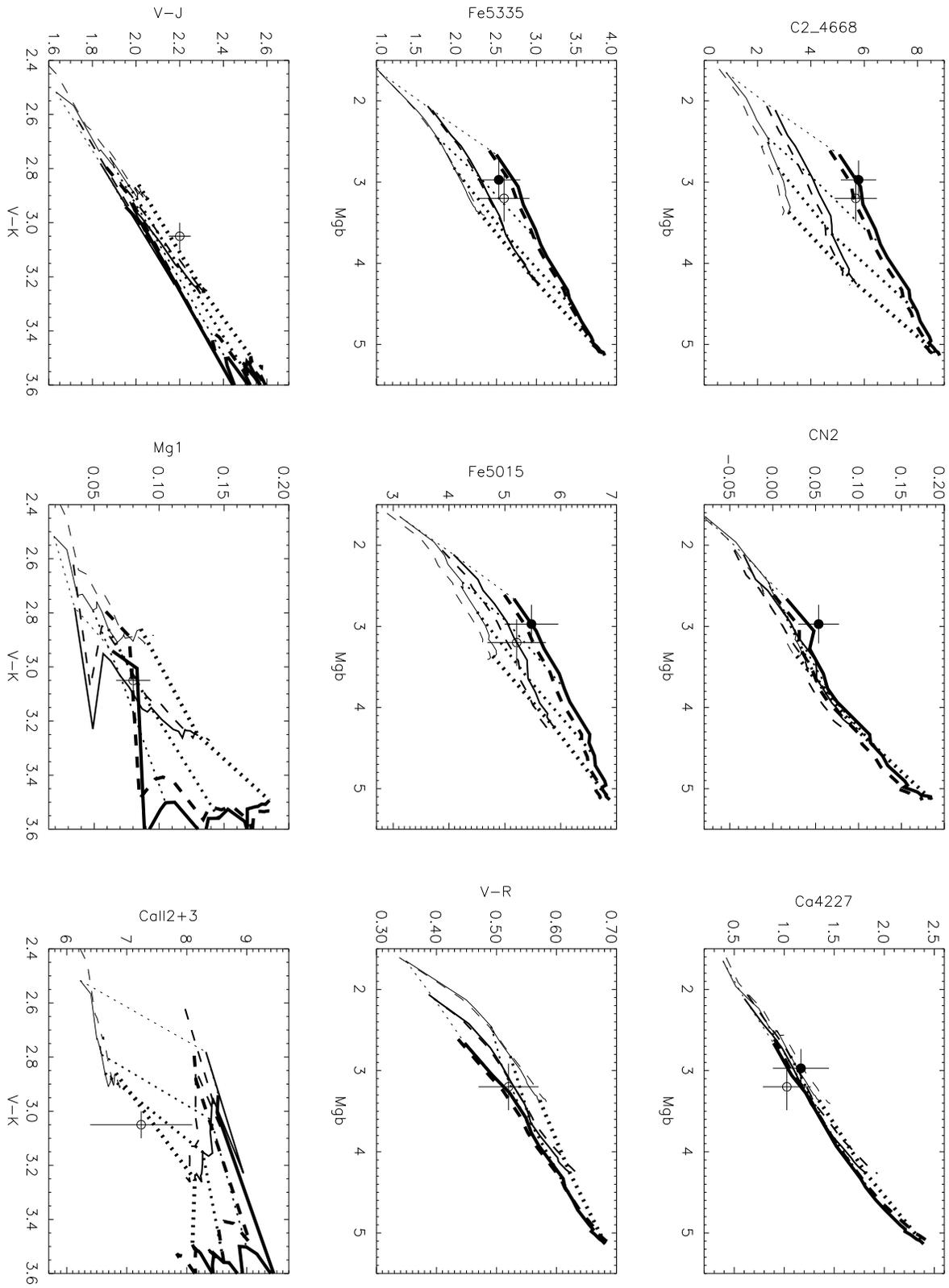}}
\caption[]{For lines and symbols see Fig.~ \ref{Figure3}.
\label{Figure4}}
\end{figure}

In Figure \ref{Figure3} and \ref{Figure4} various index-index and colour-index diagrams together
with the model predictions of V96 (for a bimodal IMF with slopes 1.35 and 2.35)
are shown. These plots show that most of the observed colours and indices indicate 
that the average metallicity of the galaxy lies between Z=0.008 and Z=0.05, and 
is very close to Z=0.02.
In addition, the spectral indices observed are between the
isolines corresponding to ages of 2.5 and 6.3 Gyrs. It is important to note
that these plots do not show any strong discrepancies between models and
observations. This allows us to conclude that element ratios in this
galaxy do not strongly differ from those in the solar neighbourhood. Although
this might seem surprising, since giant ellipticals generally have Mg/Fe
abundance ratios that are larger than solar (e.g.  Peletier
1989, Worthey et al. 1992, Vazdekis et al. 1997, Trager et al. 2000), it is in agreement with previous results, since the Mg/Fe overabundance seems to correlate with the luminosity of a galaxy, in
a way that faint ellipticals have solar abundance ratios (e.g. Worthey 1998,
Trager et al. 2000).

To be able to obtain a more detailed fit and to constrain the number of possible solutions
we used all the information provided by our large set of indices and colors
via the Merit Function MF$^J$(A,Z), which, for each model J, depends on the age
(A) and the metallicity (Z):

\begin{equation}
 MF^J(A,Z) = \frac{\sum_{i=1}^{n}~W_{i}~\times~\left(\frac{O_{i}-M_i^J}{E_{i}}\right)^2}{\sum_{i=1}^{n}~W_{i}},
\end{equation}
        
where n is the number of observed spectral indices, O$_{i}$ is the observed
value of the index i \footnote{Here the index i runs either for a color or a
spectral index.}, M$_i^J$(A,Z) is the synthetic value for this color/index
predicted by model J, and E$_i$ is its corresponding observational error.
Finally, W$_i$ represents its relative weight. Following the fact that we have
not found any significant departure from the solar neighbourhood element ratio
trends we have chosen to assign a weight of $1$ to all the indices. This
approach differs from that used in Vazdekis et al. (1997) and P99 for fitting a
sample of considerably more luminous galaxies, which showed appreciable
differences in their element ratios. These authors demonstrated that the
inferred age/metallicity depends on the relative weights of the colours and
indices (see also Vazdekis 1999a, Kuntschner 2000). For
example, they found that the analysis on the basis of the Fe dominated indices
yielded a lower metallicity than the one obtained on the basis of the Mg
dominated features. Fortunately, the solution for M~32 is to a 
large extend free from
these effects and the models (which are based on solar element ratios) seem
fully appropriate, making it possible to use the information provided by
the  whole set of indices.

We built up contour plots of MF$^J$(A,Z) for each J-model using an
interpolation program based on the Renka-Cline method described in the Numerical Recipes (Press et al. 1992). We recall that this method warrants that the reconstructed surface is  continuous with continuous first derivatives. The minimum of this surface was then determinated using the downhill simplex method carried out by the {\small AMOEBA} subroutine of the Numerical Recipes. Therefore, for each J-model we could determine the most probable solution (for the metallicity and the age) where MF$^J$(A,Z) yield the lowest values.  

We used various types of models: {\it i)} models of V96 with a
unimodal, Salpeter IMF with
slope x=1.35 (M1), {\it ii)} models of V96 with unimodal IMF and x=2.35 (M2),
{\it iii)} models with bimodal (as defined in V96) IMF  with x=1.35 (M3),
{\it iv)} models with bimodal IMF with x=2.35 (M4), and
{\it v)} W94 models with a Salpeter IMF (M5).

 \begin{table}
\footnotesize
 \caption[]{Observed and modeled spectral indices}
 \label{MODall}
 \[
 \begin{array}{p{.08\linewidth}p{.07\linewidth}cccccccccc}
 \hline
     \hline
 \noalign{\smallskip}
 \ Index & Units & Mean Obs & Error & Source & Obs - M_1 & Obs - M_2 & Obs - M_3 & Obs - M_4 & Obs - M_5 \\ 
 \noalign{\smallskip}
 \hline 
 U-V         & mag    & 1.28 & 0.05 & (P93)^{\mathrm{a}} &  -0.07 & -0.06 &  -0.06 &  -0.04 &   0.03 \\
 B-V         & mag    & 0.85 & 0.05 & (P93)		&  -0.04 & -0.05 &  -0.04 &  -0.03 &  -0.07 \\
 V-R         & mag    & 0.52 & 0.05 & (P93)		&  -0.03 & -0.05 &  -0.03 &  -0.03 &  -0.04 \\
 V-I         & mag    & 1.13 & 0.05 & (P93)		&  -0.02 & -0.09 &  -0.01 &  -0.02 &  -0.05 \\
 V-J         & mag    & 2.20 & 0.05 & (P93)		&   0.00 & -0.09 &   0.02 &   0.04 &   0.03 \\
 V-K         & mag    & 3.05 & 0.05 & (P93)		&  -0.12 & -0.18 &  -0.11 &  -0.06 &  -0.01 \\
 H$\gamma$a  & \AA & -4.1 & 0.5  & (D15)^{\mathrm{b}} &    0.5 &   0.3 &   0.5  &    0.4 &   -0.3 \\
 H$\gamma$f  & \AA & -0.7 & 0.3  & (D15)		&   -0.1 &  -0.1 &   -0.1 &   -0.1 &   -0.5 \\
 CN1         & mag    & 0.01 & 0.02 & (T98)^{\mathrm{c}} &  -0.01 &  0.01 &  -0.01 &   0.01 &   0.01 \\
 CN2         & mag    & 0.05 & 0.02 & (T98)		&   0.00 &  0.01 &   0.00 &   0.01 &   0.02 \\
 Ca4227      & \AA & 1.0  & 0.2  & (D15)		&   -0.3 &  -0.4 &   -0.3 &   -0.3 &   -0.1 \\
 G4300       & \AA & 5.7  & 0.5  & (D15)		&    0.8 &   0.9 &   0.8  &    0.9 & 	1.2 \\
 Fe4383      & \AA &  5.0 & 0.5  & (D15)		&    0.2 &   0.3 &   0.2  &    0.3 & 	0.5 \\
 Ca4455      & \AA &  1.5 & 0.3  & (D15)		&    0.0 &   0.0 &   0.0  &    0.0 & 	0.0 \\
 Fe4531      & \AA &  3.0 & 0.5  & (D15)		&   -0.3 &  -0.3 &   -0.2 &   -0.3 &   -0.2 \\
 C$_{2}$4668 & \AA &  5.7 & 0.8  & (D15)		&    0.9 &   1.2 &   0.9  &    1.2 & 	1.2 \\
 H$\beta$    & \AA &  2.0 & 0.3  & (D15)		&    0.0 &   0.1 &   0.0  &    0.0 &   -0.1 \\
 Fe5015      & \AA &  5.2 & 0.5  & (D15)		&   -0.2 &   0.0 &   -0.2 &   -0.1 &   -0.1 \\
 Mg1         & mag    & 0.08 & 0.02 & (D15)		&  -0.01 & -0.01 &  -0.01 &  -0.01 &   0.00 \\
 Mg2         & mag    & 0.20 & 0.01 & (D15)		&  -0.02 & -0.03 &  -0.02 &  -0.02 &  -0.01 \\
 Mgb         & \AA &  3.2 & 0.3  & (D15)		&   -0.2 &  -0.3 &   -0.2 &   -0.2 & 	0.0 \\
 Fe5270      & \AA &  3.2 & 0.3  & (D15)		&    0.4 &   0.4 &   0.4  &    0.4 & 	0.4 \\
 Fe5335      & \AA &  2.6 & 0.3  & (D15)		 &    0.1 &   0.1 &   0.1  &	0.1 &	 0.1 \\
 Fe5406      & \AA &  1.9 & 0.3  & (D15)		 &    0.3 &   0.3 &   0.3  &	0.3 &	 0.3 \\
 Fe5709      & \AA &  1.1 & 0.2  & (T98)		 &    0.1 &   0.2 &   0.1  &	0.2 &	 0.2 \\
 Fe5782      & \AA &  0.8 & 0.2  & (T98)		&    0.0 &   0.0 &   0.0  &    0.0 & 	0.0 \\
 NaD         & \AA &  3.4 & 0.2  & (T98)		&    0.4 &  -0.1 &   0.5  &    0.2 & 	0.6 \\
 TiO1        & mag    & 0.037& 0.007& (T98)		& -0.003 &-0.010 & -0.003 & -0.003 & -0.010 \\
 TiO2        & mag    & 0.061& 0.006& (T98)		& -0.001 &-0.014 &  0.001 &  0.000 & -0.016 \\
 CaII+       & \AA &  7.2 & 0.8  & (D15)		&   -1.2 &  -0.9 &   -1.2 &   -1.0 & 	... \\
 \noalign{\smallskip}
 \hline		    
 \end{array}	    
 \]
   \begin{list}{}{}
   \item[$^{\mathrm{a}}$] Simulated long-slit data of a region 2$^{\prime\prime}$\ from the centre (Peletier, 1993).
   \item[$^{\mathrm{b}}$] 2D$\_$FIS data corresponding to the mean values in the
    9$^{\prime\prime}$$\times$12$^{\prime\prime}$\
    central region obtained on 15 February.
   \item[$^{\mathrm{c}}$] long-slit data of the centre (T98).
 \end{list}
 \end{table}

 \begin{table}
\footnotesize
 \caption[]{Ages and metallicities for M32}
 \label{resMOD}
 \[
 \begin{array}{p{.08\linewidth}cccc}
 \hline
     \hline
 \noalign{\smallskip}
 Model &      Z      &    Age   &  MF^J(A,Z) \\
          &             &  (Gyr)   &                      \\
 \noalign{\smallskip}
 \hline 
 M1  &  0.02  &  4.0  &  1.0  \\
 M2  &  0.02  &  4.0  &  1.8  \\
 M3  &  0.02  &  4.0  &  1.0  \\
 M4  &  0.02  &  4.0  &  0.8  \\
 M5  &  0.019 &  5.0  &  1.4  \\
 \noalign{\smallskip}
 \hline		    
 \end{array}	    
 \]   
 \end{table}

\section{Results}

\subsection{A single age-metallicity solution}

\begin{figure}
\mbox{\epsfysize=21cm  \epsfbox{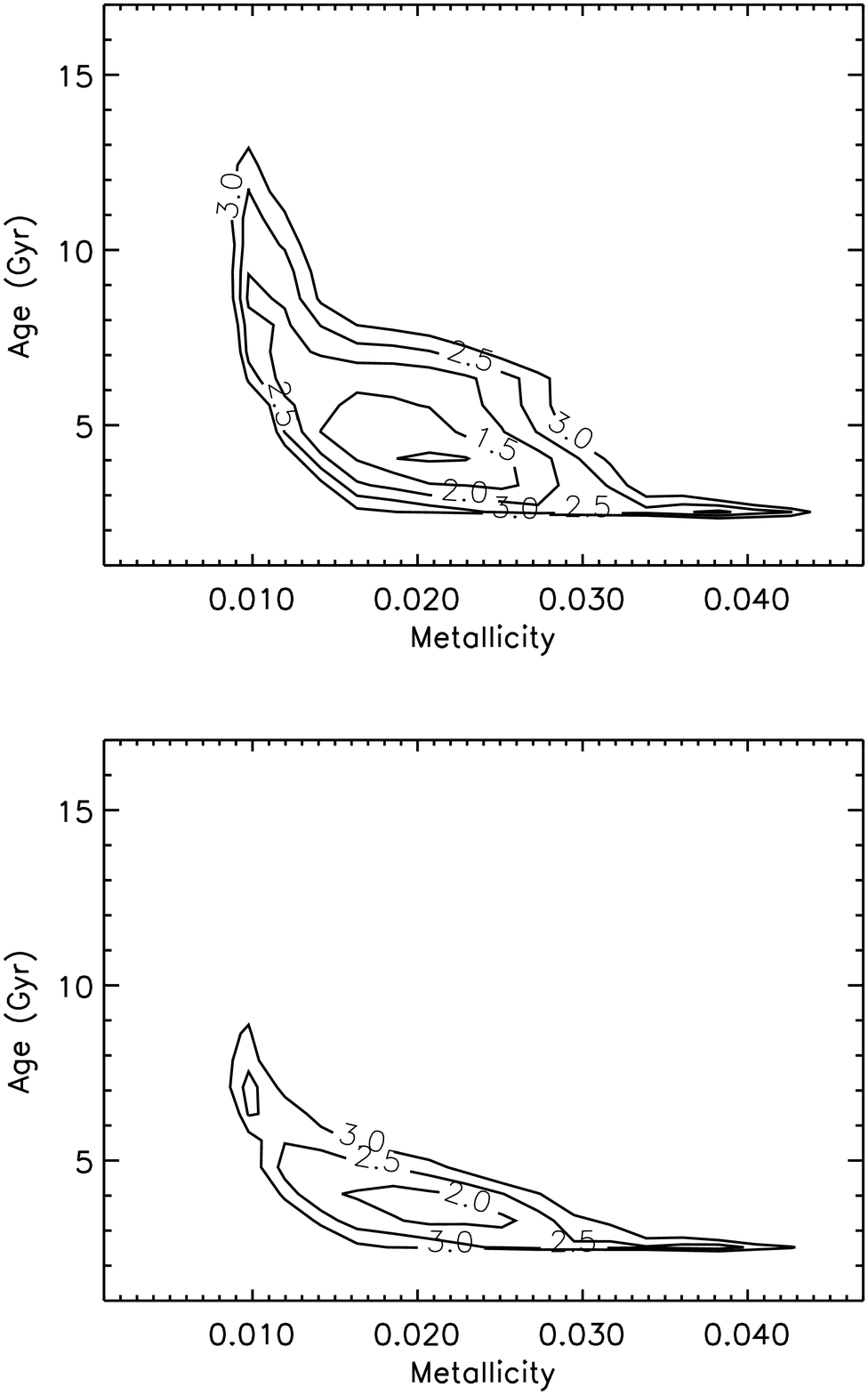}}
\caption[Figure5.ps]{Contours of the 
Merit Function for unimodal models with IMF slope x=1.35 (top)
and x=2.35 (bottom). \label{Figure5}}
\end{figure}

\begin{figure}
\mbox{\epsfysize=21cm  \epsfbox{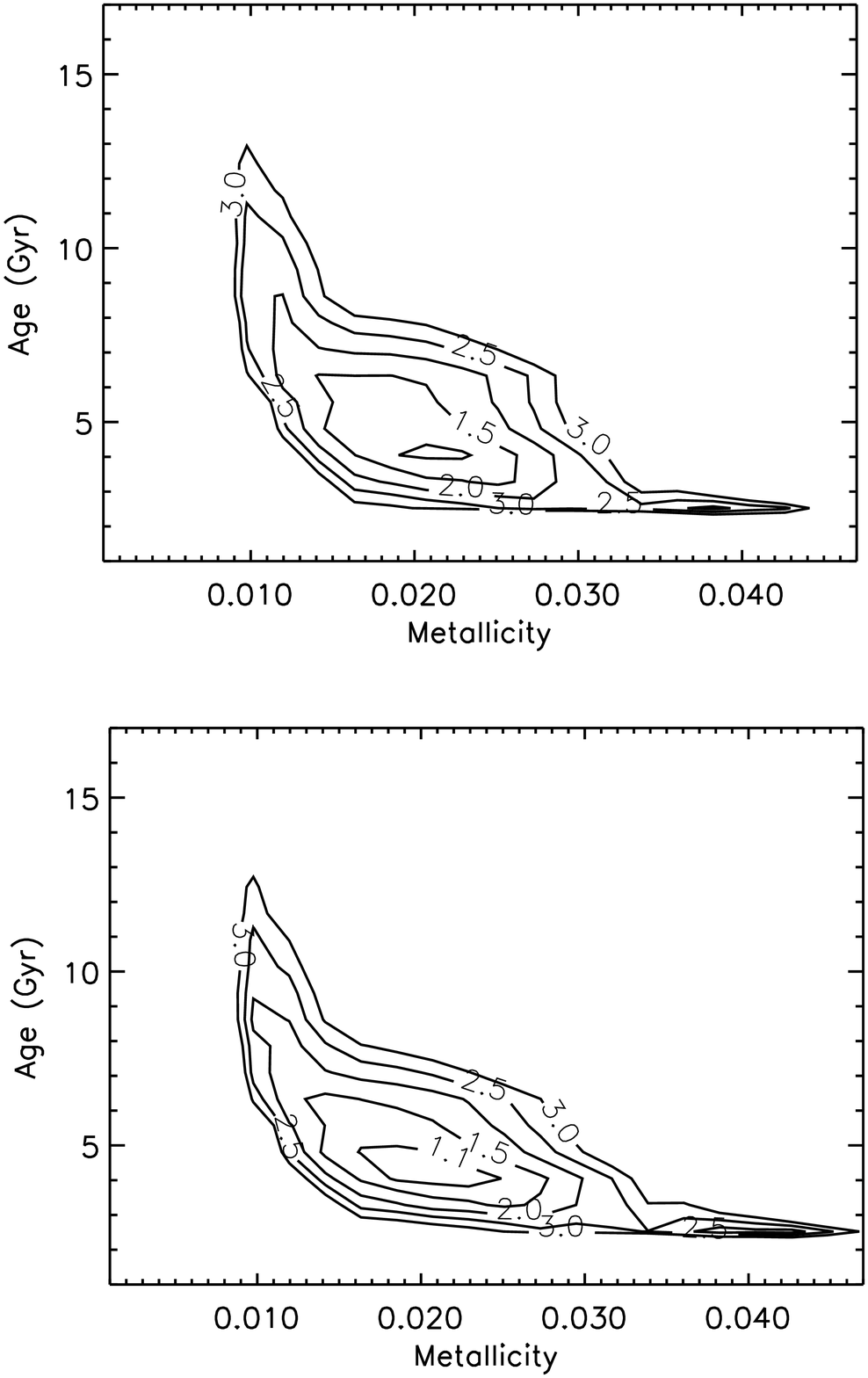}}
\caption[Figure6.ps]{Contours of the Merit Function for bimodal models with IMF slope x=1.35 (top) and
x=2.35 (bottom). \label{Figure6}}
\end{figure}

\begin{figure}
\mbox{\epsfxsize=14cm  \epsfbox{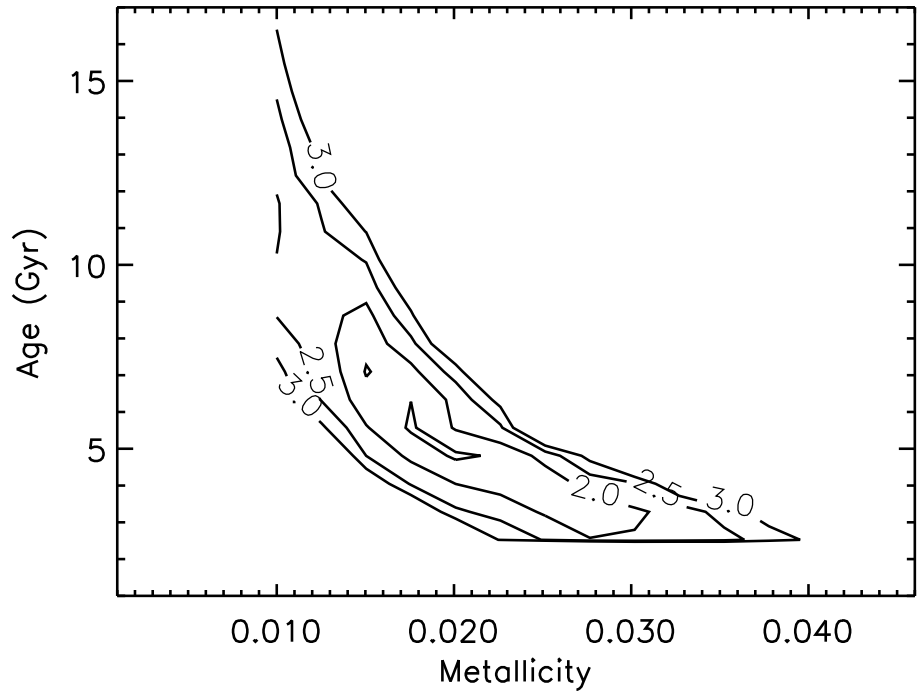}}
\caption[Figure7.ps]{Contours of the Merit Function for the Worthey model. 

\label{Figure7}}
\end{figure}

Figs. \ref{Figure5}, \ref{Figure6} and \ref{Figure7} show contour plots of the
MF as a function of metallicity (Z) and age (A) for the different models. MF
values lower than 1 represent fits which are fully acceptable within the error
bars of the observed data. However, to prevent the exclusion of any possible
solution we consider a fit acceptable when MF value is lower than
1.5. In Table \ref{resMOD} we tabulate the best fits for the age and the
metallicity and the corresponding MF$^J$(A,Z) values achieved with each model.
We see that all the models predict a similar age ($\sim$4~Gyr) and a solar-like
metallicity (W94 models provide a slightly larger age $\sim$5~Gyr). The
contours of the MF plots also show that fits with ages in the range (3-6~Gyr)
and metallicities in the range (0.015-0.025) cannot be discarded. It is worth
recalling that the contours of the MF are indicating the well known
age-metallicity degeneracy: older ages require lower metallicities and
viceversa. However, the contours are more elongated toward the older ages than to the younger ones because the indices and colors of the stellar populations evolve more slowly for larger ages (e.g., W94; V96). It is also interesting to see that these contours are more elongated for the W94 models. The MF plots and Table \ref{resMOD} show some differences between the models. In fact, M2 does not provide any reasonable fit to the observed indices and colors. Overall, V96's models with an unimodal IMF and slope x=1.35 (M1) or the models with bimodal IMFs (M3,M4) provide the best fits. The best solution is
reached when using M4 (MF=0.8). Table \ref{MODall} lists the differences between observations and models (M1, M2, M3, M4 and M5) for colors and indices corresponding to each solution (A,Z) shown in Table \ref{resMOD}. In this table MF$^J$(A,Z) for each solution (A,Z) is also listed.        

\subsection{Giants versus dwarfs}

In Table \ref{DwaGia} we tabulate the relative contribution of the dwarfs 
to several spectral bands from ultraviolet to infrared. We do this for SSPs of
4~Gyr and solar  metallicity, using V96's models. We see that the dwarfs 
contribute $\sim$80\% of the total light in the U-band, while in K the giants
dominate the total light with a very similar proportion. In V, the dwarfs and
giants contribute $\sim$50\% each.

 \begin{table}
\footnotesize
 \caption[]{Contribution of the dwarfs (in \%) to the total light for a 
 SSP of 4~Gyr and Z=0.02}
 \label{DwaGia}
 \[
 \begin{array}{p{.07\linewidth}ccccccccc}
 \hline
     \hline
 \noalign{\smallskip}
 Model & U & B & V & R & I & J & H & K \\
 \noalign{\smallskip}
 \hline 
 M1 & 80.1 & 61.7 & 49.0 & 42.8 & 35.2 & 22.4 & 16.7 & 14.6 \\
 M2 & 83.1 & 67.2 & 56.6 & 52.3 & 48.9 & 40.3 & 34.1 & 32.3 \\
 M3 & 80.1 & 61.7 & 49.2 & 43.1 & 36.0 & 23.8 & 18.0 & 15.9 \\
 M4 & 83.0 & 66.7 & 55.4 & 50.1 & 43.6 & 31.0 & 24.7 & 22.2 \\
 \noalign{\smallskip}
 \hline		    
 \end{array}	    
 \]    
 \end{table}

Table \ref{DwaGia} suggests that the changes to colors as a result of
varying the IMF are largest for the largest wavelength range between
the 2 bands constituting this colour.  This means that
the effect on a color like $V-K$ is much larger than e.g. $V-R$  (in agreement
with Vazdekis 1999b). It should be noted, however, that the effects of changing
the IMF, for the bimodal case, are much smaller than for the unimodal case,
since in the bimodal case there is a lower number of  stars with
masses smaller than 0.6 M$_\odot$, the stars which are red enough  to
contribute substantially to the $V-K$ color. The table also shows that the
relative contribution of dwarfs and giants does not vary very much from model
to model, except for model M2, which is unimodal, with x=2.35. Even here
significant differences are only seen for the reddest bands.  In fact, this
model is not able to fit the galaxy because it provides near-IR colors (e.g.,
$V-K$) that are too red (see Table \ref{MODall}).  Note, however, that the
optical line-strengths of this model (with the exception of the reddest ones, 
like the  TiO-bands) are not very different from those of the models M1, M3 and
M4. M1, M3 and M4 give similar fits, although model M4, which has a
slightly larger contribution of dwarfs than M1 and M3, provides a somewhat 
lower MF value. 

\section{Discussion}

\subsection{Gradients}

The optical and infrared photometry shows no evidence for the presence of gradients in the colors (Peletier 1993, Lauer 1998). The V-I and U-V images ofthe HST (Lauer, 1998) are essentially flat, without any sign of structure (dust, inner disk, etc.). Ultraviolet colours show strong gradients, e.g., in the (152 -- 249) colour (O'Connell et al. 1992) and the m$_{1500}$ -- B colour (Ohl et al. 1998), in the sense that M32 is redder in the centre. Its behaviour is opposite to all other objects studied by these authors, and also therefore not well understood. It is not in conflict with the optical data, since a few hot stars can affect the UV bands without noticeably altering the optical colours. A possible explanation could be that the m$_{1500}$ -- B gradient is caused by a slightly older centre.

Concerning the line-strength gradients, Gonz\'alez (1993) pointed out that
various indices show relatively small, but monotonically decreasing  profiles,
particularly H$\beta$ and Fe5270. Davidge (1991) detected  a very small
gradient in the inner 5~arcsec. Hardy et al. (1994) noted  that the line
indices are smaller in the outer parts. However, our  2D\_FIS data do not show
any significant gradient for the central 
9$^{\prime\prime}\times$12$^{\prime\prime}$ region of this galaxy. Also, the data of T98 rule out any variation of the line strength profiles in the inner 
15~arcsec.  The H$\beta$ gradient of Gonz\'alez (1993) suggests that the galaxy 
is younger in the centre. However, to obtain a  constant color or Mg$_2$
profile the galaxy needs to be at the same time more metal-rich towards the
centre, so that both effects cancel each  other out. In the inner areas
discussed here these gradients are negligible. Measurable age- or metallicity gradients
could be present when going to larger radii, but this is outside the scope of
this paper.

\subsection{Age and Metallicity}

The large number of indices and colors used in this work allowed us  to
constrain the age and the metallicity  for this galaxy, although  the
contours of the MF plots still show some of the effects of the age-metallicity
degeneracy. The best fits suggest an age of 4~Gyr and a solar-like metallicity. This
result is in agreement with previous results (O'Connell 1980; Burstein et al. 
1984; Rose 1985,1994; Faber et al. 1992; Hardy et al. 1994; Dorman et al. 1995;
Vazdekis \& Arimoto 1999). The spectrophotometric study of O'Connell (1980) 
suggested that the turnoff stars should be close to F8.  Burstein et al. (1984)
also conclude that turnoff stars with spectral types F6--F9 are needed to
make H$\beta$ strong enough to fit the value for M~32. In that paper the authors
rule out models with many blue stragglers or a strong BHB.  
Dorman et al. (1995)  exclude the
presence of a representative contribution of RHB stars and conclude  that M~32
has an intermediate age and a solar-like metallicity. Boulade et al.  (1988)
reach the same conclusion from the spectral
range between $\lambda\lambda$3200 and 3900~\AA~.  
Rose (1985, 1994) on the basis of the spectral  range around
$\lambda$4000\AA~concluded that this galaxy has no significant contribution
from a RHB population, and has a more metal-rich main-sequence  turnoff and a
larger hot star contribution than is inferred to be present in  47~Tuc. More
recently Rose \& Deng (1999) reach the same conclusions on the basis of an 
analysis of IUE
mid-ultraviolet low resolution spectra. Our result is in full
agreement with that of Vazdekis \& Arimoto (1999), who predicted an age of 4~Gyr and
solar-like metallicity from a very good  fit to the blue spectrum of this
galaxy, using a new spectroscopic age indicator.  

There are, however, other authors who suggest a larger age. Jones \& Worthey 
(1995) obtained 7~Gyr based on their H$\gamma_{HR}$ index.  Vazdekis \& Arimoto
(1999) attributed this difference to the fact that W94  model makes use of
isochrones which are slightly hotter than those of  V96's model and to the
fact that H$\gamma_{HR}$ is very sensitive to the  resolution and to the velocity
dispersion effect. A relatively recent HST  analysis resolving M~32 into
individual stars (Grillmair et al. 1996) suggests a wide metallicity spread
(but smaller than expected from the  {\it simple model}) and the presence of
SMR stars on the  basis of the observed color spread of AGB and RGB stars. 
Also, their
color-magnitude diagrams do not indicate the presence of bright AGB stars as
claimed by  Freedman et al. (1992).  Although they do not distinguish between
age and metallicity using  colors, they infer an age of 8.5 Gyr (for at least a
50\% of its stellar  populations) and a metallicity [Fe/H]=-0.25 (Z=0.014). Although their  solution for the age and metallicity is inferred
from the indices of  T98 using W94 models, our analysis (see model M5 in  Table
\ref{resMOD}) is in disagreement with their conclusion. In our opinion their data suffers 
from the age-metallicity degeneracy, and resolving individual stars
does not help in solving this problem.

\subsection{Unique solution}

Our data have been successfully fitted with a single-age, single-metallicity
stellar population as has been done by 
other authors (e.g.: Vazdekis \& Arimoto 1999). However, it has been suggested
that M~32 contains 2 stellar populations, an underlying old stellar population,
together with a second stellar population of younger stars (see, e.g., O'Connell 1980). We have tested this idea, and investigated what the best solution is, for a linear combination consisting of an old population of 17 Gyr and Z=0.02,
and a young population with Z=0.02 and variable age between 1 and 5 Gyr, 
using model M4. The results are given in Table 7.
As expected, the relative contribution of the young stellar
population decreases when
considering ages closer to 1 Gyr. We find that the 
minimum MF is obtained for a combined
stellar population of 3.2 Gyrs and 17 Gyrs (82\% and 18\%, respectively) but
all the tabulated combinations are acceptable, because the 
MF$<$1.5. To really constrain the 2-population model, more indices in the 
near and far UV are needed.

 \begin{table}
\footnotesize
 \caption[]{Percentage of the young stars with ages between 1-4 Gyrs for a combined stellar population formed by these stars and others with 17 Gyrs (fixed Z=0.02) according to model M4. The Merit Function values are also tabulated.}
 \label{Ratio}
 \[
 \begin{array}{p{.15\linewidth}ccccccccc}
 \hline
     \hline
 \noalign{\smallskip}
 Age (Gyrs)    & 1.0  &  1.6  &  2.0  &  2.5  &  3.2  & 4.0  & 5.0 \\
 \noalign{\smallskip}
 \hline 
 \% Young pop. &  35  &   50  &  58   &   69  &   82  & 100. & 100. \\
 MF            & 1.32 & 1.12  & 1.02  &  0.91 &  0.88 & 0.94 & 1.34 \\
 \noalign{\smallskip}
 \hline		    
 \end{array}	    
 \]    
 \end{table}

\subsection{Abundance Ratios}

Our 2D\_FIS data do not show any significant departure from solar-like element
ratios. However, we have found a number of indices which slightly differ from the models, such as C$_{2}$4668, which is slightly larger than the predicted
values. This feature has been claimed by Worthey (1994) as a strong metallicity
indicator, and Tripicco \& Bell (1995) showed  that this is an
indicator of the global
metallicity. Kuntschner \& Davies (1998) found that the H$\gamma$A-C$_{2}$4668
diagram is very useful for studying the age and the metallicity. 
More recently  Kuntschner (2000) suggested that this index
seems to be too strong compared to Fe for galaxies for which the C$_{2}$4668
line-strength is larger than $\sim$6 \AA. We obtain 5.7 \AA\ while a slightly
higher value is tabulated in T98. However, we should be aware that this index
has a wide central bandpass (86.25~\AA) and therefore could be affected by
systematic errors. Note for example the large transformation factors given
in Table 3 of P99. The G band is significantly larger than the model
predictions but this difference is not as large if we consider the value of
T98. In fact, we do not exclude that some systematic error in this spectral 
range could be affecting our data here, as well. The Ca{\sc II} triplet feature is low
compared with the solar metallicity model predictions. It is surprising that
our value (7.2 \AA) is
very similar to the  mean value of the nuclei of normal galaxies (7.1 \AA,
Terlevich et  al. 1990) and also the three galaxies in P99 that were observed
with the same instrument, even though M~32 is much less massive, and 
other indices are generally much smaller. This indicates that the Ca triplet
could be a great giant/dwarf discriminator.
However, we point out that the model predictions for
this index are much less accurate than the others 
because of the lack of completeness of the library of D\'{\i}az et al. (1989). 
A new extensive stellar spectral library covering this feature will be soon available (Cenarro et al. 2000, in preparation).

Our results provide reasonably good fits to the other absorption
line-strengths. It is interesting to note that we obtain an excellent fit to
the strength  of H$\beta$ while O'Connell (1980) found that it was 3$\sigma$
too weak  compared to his models. The difference is probably due to the  fact
that the models are now better than before.

For most of the indices the obtained fits provide line-strengths that fall 
within the error bars of the measurements (i.e., MF$\sim$1). However,  there
are a number of indices which deviate slightly from the predicted values,
in a systematic way. For
example,  for the Ca{\sc II} triplet features the observed values are lower
than the model predictions for solar metallicity (see also P99). The predicted values,
however, might not be very accurate (see the above discussion). The Ca4227 line
is also slightly lower than the model predictions, but, on the other hand, the
value  provided by T98 is slightly higher
than ours. We also see that the G band it is 1.8 $\sigma$ larger than in the
models. This difference, however, is reduced if we use instead the G-band 
value given
by T98, which is considerably lower. This suggests that our G band measurement
could be affected by fringing in the 2D\_FIS spectra. Concerning the Mg/Fe
relation we see that all the residuals for these lines are within the
error bars. Note, however,  that while the residuals of the Mg lines are
negative most of the residuals of the Fe features are positive, except for
indices Fe4531 and Fe5015.

\subsection{Dwarfs/Giants} 

The best fits obtained here suggest that the dwarfs contribute to the total
light $\sim$80-83\% in the U filter, $\sim$61-67\% in B, $\sim$49-56\% in V,
$\sim$43-50\% in R, $\sim$35-44\% in I, $\sim$22-31\% in J, $\sim$17-25\% in H
and $\sim$14-23\% in K. These values have been obtained on the basis of
models M1, M3 and M4  (i.e., V96 models with  unimodal IMF with slope 1.3 and
bimodal models with slope 1.3 and 2.3). Rose (1994) used various diagrams such
as SrII$\lambda$4077/FeI$\lambda$4045 versus H$\delta$/FeI$\lambda$4045 to
conclude a dwarfs/giants relation of 0.65:0.35 which is in perfect agreement
with our estimate for the B filter. The luminosity-weighted mean surface
gravity for M~32 at 3500 \AA\ indicates that the giant branch contributes
slightly less than 30\% of the light at 4000 \AA\ (Boulade et al. 1988). This
is in agreement with our estimation in the filter U.

Concerning the IMF slope, CO (2.3$\mu$m) data of early-type galaxies  (not
including M~32) with -19$\le M_v \le$-23 could be adequately fitted  by models
with IMF slope x=1.35 (Aaronson 1978). He gave an upper limit for  x=2.25. Here we have
shown that the main changes introduced by changes in the IMF should be found
for the spectral indices and colors centred at the longest wavelengths. In
fact, we see significant differences in, e.g., the V-K color or the Ca{\sc
II} features. However, the present model and observational uncertainties do not
allow us to constrain the IMF more strongly.

\section{Conclusions}

We have obtained Integral Field Spectroscopy of the 9$^{\prime\prime}\times$12$^{\prime\prime}$ circumnuclear region of M32. From these spectra, we have applied a stellar population analysis using 30 spectral indices and colours. Our main conclusions are:

\begin{enumerate} 
\item None of the spectral index maps show significant
variations in the sampled area. This result is in agreement with long-slit
spectroscopy in the literature (T98).
\item From the population synthesis models of V96 and
W94 we find that the best-fitting models have an age of 4 Gyrs 
and a solar-like metallicity for the
nuclear region of M~32. We have found that the element abundance ratios
are generally the same as in the sun.  Models with a Salpeter IMF with
a slope of 2.35 or larger are excluded. Although our data can be successfully interpreted with a single age-metallicity
stellar population, it is also possible to fit them with a linear
combination of an old stellar population of 17 Gyr and Z=0.02 and a young
stellar population with age 1-5 Gyr.

\item Our best fits using the model of V96 suggest that the
dwarfs contribute to the total light $\sim$80-83\% in the U filter,
$\sim$61-67\% in B, $\sim$49-56\% in V, $\sim$43-50\% in R, $\sim$35-44\% in I,
$\sim$22-31\% in J, $\sim$17-25\% in H and $\sim$14-23\% in K.

\end{enumerate}

\section*{Acknowledgements}

This paper is based in part on observations obtained at the William Herschel
Telescope, which is operated at La Palma by the Isaac Newton Group in the Spanish Observatorio del Roque de los Muchachos of the Instituto  de Astrof\'\i sica de Canarias. We would like to thank the anonymous referee for their suggestions. We are grateful to Adolfo Garc\'\i a for his work developing 2D-FIS and Luis Cuesta for the code GRAFICOS. We thank B. Garc\'\i a, C. Guti\'errez, and F. Prada for help with the observations and planning this project. This work has been partially supported by the Direcci\'on General de Investigaci\'on Cient\'\i fica y T\'ecnica (PB93-0658).

\newpage

\section*{Figure Captions}

\end{document}